\documentclass[aps,pra,floatfix,twocolumn,superscriptaddress]{revtex4}

\pdfoutput=1
\usepackage{graphicx}
\usepackage{amssymb}
\usepackage{amsmath}

\newcommand {\ket}[1] {|#1 \rangle}
\newcommand {\bra}[1] {\langle#1 |}

\begin{document}

\title{State-dependent lattices for quantum computing with alkaline-earth-metal atoms}
\author{Andrew J. Daley}
\affiliation{Institute for Quantum Optics and Quantum Information of the
Austrian Academy of Sciences, A-6020 Innsbruck, Austria} \affiliation{Institute
for Theoretical Physics, University of Innsbruck, A-6020 Innsbruck, Austria}
\affiliation{Department of Physics and Astronomy, University of Pittsburgh, Pittsburgh, PA 15260, USA}

\author{Jun Ye}
\affiliation{JILA, National Institute of Standards and Technology and University of Colorado\\
and Department of Physics, University of Colorado, Boulder, CO 80309-0440, USA}

\author{Peter Zoller}
\affiliation{Institute for Quantum Optics and Quantum Information of the
Austrian Academy of Sciences, A-6020 Innsbruck, Austria} \affiliation{Institute
for Theoretical Physics, University of Innsbruck, A-6020 Innsbruck, Austria}

\date{\today}

\begin{abstract}
Recent experimental progress with Alkaline-Earth atoms has opened the door to quantum computing schemes in which qubits are encoded in long-lived nuclear spin states, and the metastable electronic states of these species are used for manipulation and readout of the qubits. Here we discuss a variant of these schemes, in which gate operations are performed in \emph{nuclear-spin-dependent} optical lattices, formed by near-resonant coupling to the metastable excited state. This provides an alternative to a previous scheme [A. J. Daley, M. M. Boyd, J. Ye, and P. Zoller, Phys. Rev. Lett \textbf{101}, 170504 (2008)], which involved independent lattices for different \emph{electronic} states. As in the previous case, we show how existing ideas for quantum computing with Alkali atoms such as entanglement via controlled collisions can be freed from important technical restrictions. We also provide additional details on the use of collisional losses from metastable states to perform gate operations via a lossy blockade mechanism.
\end{abstract}

\maketitle

\section{Introduction}

There has been a lot of recent experimental progress in cooling and manipulating alkaline-earth and alkaline-earth-like atoms in the laboratory, especially in the context of optical clocks with Strontium Atoms  \cite{AEatom1,zeemanshift1,Boyd07,ludlow08,Boydthesis}, and the production of Bose-Einstein condensates and degenerate Fermi gases of Ytterbium \cite{takasu03,takasu07,Takahashi}, Calcium \cite{Kraft2009} and Strontium \cite{Stellmer2009,deEscobar2009,DeSalvo2010,Tey2010dgb}. The control that has been developed over these atoms makes them an extremely interesting candidate for the implementation of quantum information processing \cite{aeshort,Gorshkov2009,hayes07,reichenbach07}. This is especially true in light of the laser stability achieved in optical clock experiments \cite{Boyd07,Boydthesis}, which is reminiscent of the development path towards quantum computing taken in the case of trapped ions \cite{ions,ions2}.

\begin{figure}[tb]
\includegraphics[width=8.5cm]{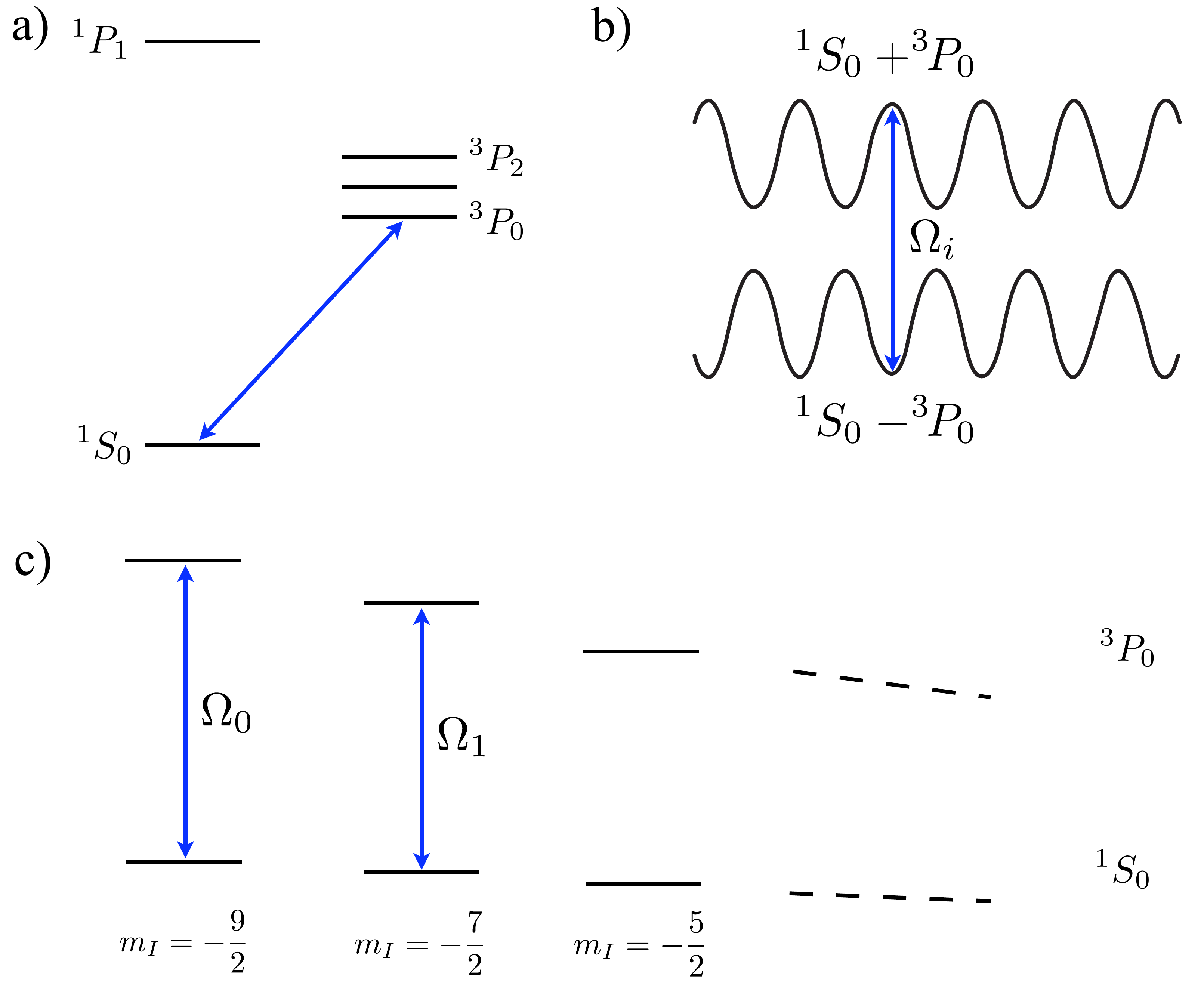}
\caption{Level structure for Alkaline-earth-like atoms. (a) These atoms possess a singlet-triplet transition with long-lived metastable $^3$P$_0$ and $^3$P$_2$ levels. (b) Adiabatic dressed potentials can be created by a resonant coupling on the clock transition with a sinusoidally varying Rabi frequency $\Omega_i(x)$, producing dressed states of the $^1$S$_0$ and $^3$P$_0$ levels. (c)  The differential Zeeman shift for different nuclear spin levels can be used to produce nuclear spin-dependent lattices, by driving the transition between the $^1$S$_0$ and $^3$P$_0$ levels resonantly at the different resonant frequencies for different nuclear spin states in a magnetic field. Here we show some of the $m_I$ states for an atom with nuclear spin $I=9/2$, such as $^{87}$Sr.} \label{fig:levelstructure}
\end{figure}

The key new feature of alkaline earth atoms in comparison with alkali atoms is the singlet-triplet metastable transition, with the $^1$S$_0$ -- $^3$P$_0$ transition being used as the clock transition (see Fig.~\ref{fig:levelstructure}a). In particular, for $^{87}$Sr, the $^3$P$_0$ manifold has a measured lifetime of $\tau \sim 30$s, and the $^3$P$_2$ levels have even longer predicted lifetimes. In addition, for species with non-zero nuclear spin, this spin can be decoupled from the electronic state on the clock transition \cite{hayes07, reichenbach07, yi08, aeshort, Gorshkov2009}, especially in the presence of a large magnetic field. The use of this nuclear spin for storage of quantum information would then be ideal, as the nuclear spin is much less sensitive to magnetic fields than electron spins, and thus much less susceptible to decoherence from magnetic field fluctuations than qubits stored on electronic states. This has lead to a series of proposals \cite{aeshort,Gorshkov2009,Shibata} in which the electronic state is used for access to and manipulation of the qubit \cite{dereviankoaddressing}, and the nuclear spin state is used for qubit storage.

In previous work \cite{aeshort} we developed a scheme for quantum computing with alkaline earth atoms that was based on \emph{electronic-state-dependent} lattices in which independent control over lattices for the metastable excited $^3$P$_0$ and ground $^1$S$_0$ levels is obtained by using light of different wavelengths. This is made possible by the fact that these levels are optically separated, providing very different AC polarisabilities for the states as a function of the wavelength. We showed how these two independent lattices could be used as a \emph{storage} lattice for qubits encoded on the nuclear spin state, and a \emph{transport} lattice to manipulate the qubits and perform gate operations \cite{aeshort}. A key theme in this context is that many schemes and concepts developed for alkali atoms, including certain techniques that have already been demonstrated in proof-of-principle alkali experiments simply work quantitatively better for alkaline earth atoms, where they are freed from important technical restrictions. In this sense, alkaline earth atoms represent an extremely important technological advance in various settings.

Here we present an alternative scheme to this previous proposal, in which we show that near-resonant coupling on the clock transition can produce frequency selective \emph{nuclear-spin-dependent} lattices (see Fig.~\ref{fig:levelstructure}b,c). As in the case of electronic-state-dependent lattices, this makes it possible to produce state-dependent lattices without the restriction of having to tune couplings between fine-structure states \cite{spinlattice1,spinlattice2}, which can lead to large heating and decoherence from spontaneous emissions in the case of alkali atoms. In addition, spin-dependent lattices made in this way can be easily generated so that motion of the two potentials is independent in 2D. Below we discuss this implementation in detail, developing a complete proposal for quantum computing with alkaline-earth(-like) atoms, including methods for production of a quantum register and for performing single-qubit operations. Qubit readout with individual addressing can be performed in a similar manner to the case of electronic-state-dependent lattices \cite{aeshort}, using magnetic gradient fields to shift the energy of states in the $^3$P$_2$ manifold. In nuclear-spin-dependent lattices, the large two-body loss rates from metastable $^3$P$_2$ levels can also be used to perform two-qubit gate via a lossy blockade mechanism as an alternative to the implementation of gates via controlled collisions \cite{spinlattice1}. This was originally discussed for electron-state-dependent lattices \cite{aeshort}, and we provide further details of this mechanism below.

This alternative scheme with nuclear-spin-dependent lattices has the advantage of not requiring additional lasers to trap the $^1$S$_0$ and $^3$P$_0$ lattices independently, and can be performed with a strong laser on the clock transition. At the same time, this method is somewhat sensitive to collisional losses when two atoms in the $^3$P$_0$ manifold collide, as the near-resonant lattices will always produce admixtures of this state. However, this is only an issue during the short times involved in gate operations, and is strongly suppressed in a realistic setup where atoms are also AC-Stark shifted, as discussed below. Nuclear-spin-dependent lattices would also have immediate applications in quantum simulation \cite{Cazalilla,Gorshkov,Hermele2009,FossFeig,Gerbier2010}. In particular, the degeneracy in models with SU(N) symmetry \cite{Cazalilla,Gorshkov,Hermele2009} (which can be studied using alkaline-earth-like atoms by making use of the symmetry for interactions of atoms in different nuclear spin levels) could be deliberately broken and restored by applying these nuclear-spin-dependent potentials. 

The rest of this article is organised as follows: We first discuss the formation of near-resonant spin-dependent optical lattices in more detail in Sec II, together with preparation of atomic registers in these lattices. In Sec.~III we then discuss means for readout of individual qubits, and in Sec IV we treat ideas for gate schemes to entangle two qubits, including making use of lossy blockade mechanisms. In Sec.~V we present a summary and outlook.
 
\section{Spin-dependent Adiabatic potentials}
\label{sec:potentials}

In the following, we discuss a quantum register formed by one atom trapped every site of a deep optical lattice, where tunnelling of atoms between sites can be neglected on the timescale of the experiment. As discussed above, we identify hyperfine states with two chosen nuclear spin states, and we would like to create spin-dependent potentials in order to move these qubit states independently. We will make use of these in the two-qubit gate operations that we discuss in Sec. IV. 

A novel method of forming optical lattices for alkaline earth atoms is to make use of a near-resonant optical coupling directly on the clock transition, which will produce adiabatic dressed potentials \footnote{For a discussion of the use of near-resonant lattices to produce polarisation-dependent potentials, see Ref.~\cite{yi08}.}. In the case that the coupling field is a standing wave, the Rabi frequency, and thus the final dressed potential, will be sinusoidally varying, providing an optical lattice for dressed states that are superposition of states in the $^1$S$_0$ and $^3$P$_0$ levels (as shown in Fig.~\ref{fig:levelstructure}b). 
In a large magnetic field, there is a differential Zeeman shift $\Delta E_Z$ between the $^1$S$_0$ and $^3$P$_0$ states (109 Hz/G for $^{87}$Sr \cite{Boydthesis}), meaning that a direct coupling preserving the nuclear spin (with $\pi$-polarised light) will be resonant at substantially different frequencies for different nuclear spin states (see Fig.~\ref{fig:levelstructure}c). We can then drive each transition independently with Rabi frequencies $\Omega_0=\Omega_{\pm}$, as shown in Fig.~\ref{fig:levelstructure}b. Provided that the shift $\Delta E_Z \gg \Omega_{\pm}$, we will then obtain independent two-level systems for each $m_I$ state for which we apply the appropriate coupling frequency. For example, if we choose $\Omega_0\sim 100$ kHz, then for $^{87}$Sr, we would like to apply a field $\gtrsim 1000$ G in order to obtain shifts between neighbouring states $\gtrsim 100$ kHz (however, states separated further in $m_I$ could also be used to reduce the required field - see below). In this way, we can choose, e.g., two $m_I$ states as our two qubit states, $\ket{0}$ and $\ket{1}$, and create independent potentials for these two states. At the same time, because the frequency differences between lattices for different $m_I$ states will of the order of 1 MHz, the lattice laser wavenumber $k_l$ is approximately the same for the two species - in fact the resulting lattice potentials will overlap for the order of millions of periods. 

\subsection{Dressed potentials for a two-level system}
We will now discuss the form of the dressed potentials for a single nuclear spin state, identified with qubit state $i=0$ or $i=1$, and discuss the case where we have multiple nuclear spin states below.
We can first write the Hamiltonian for a two-state atom, with states $\ket{g,i}\equiv\ket{^1{\rm S}_0,m_I=i}$ and $\ket{e,i}\equiv\ket{^3{\rm P}_0,i}$ as ($\hbar\equiv 1$) 
 \[\hat H=\hat H_M+\hat
H_0,\] where $\hat{H}_M=\hat{\mathbf{p}}^2/2m$ is the 
kinetic energy, and \[\hat H_0=-\delta_i\ket{e,i}\bra{e,i}+(\Omega_i(\mathbf{x})/2)\ket{e,i}\bra{g,i}+{\rm h.c.} \] describes the near-resonant coupling field with $\Omega_i(\mathbf{x})$ and $\delta_i$ the Rabi frequency and detuning respectively.

Generation of adiabatic dressed potentials is then based on the validity of a Born-Oppenheimer-type assumption, where we assume that the kinetic energy of the atoms is small on a scale given by the separation of the resulting adiabatic potentials. The wavefunction
$|\Phi(t)\rangle$ of a single atom satisfies the Schr\"odinger
equation,
\begin{equation}
i\hbar\frac{\partial}{\partial t}|\Phi(t)\rangle=(H_M+H_0)|\Phi(t)\rangle.
\end{equation}
If we omit the kinetic energy term from the Hamiltonian, we obtain an equation for adiabatic eigenstates, $\ket{\Psi_{\pm}(t)}$,
\begin{equation}
H_0(\mathbf{x})|\Psi_{\pm}\rangle=V^{\pm}(\mathbf{x})|\Psi_{\pm}\rangle,
\end{equation}
Note here that as $H_0(\mathbf{x})$ is time-independent, there are only two such eigenstates $\ket{\Psi_{\pm}}$. If we consider the 1D case, and set $\Omega_i(x)=\Omega_i \sin(k_l x + \phi)$, representing the field of a standing wave (with $k_l$ the laser wavenumber and $\phi$ a phase), we find the adiabatic potentials $V^{\pm}({x})=(-\delta_i\pm\sqrt{\delta_i^{2}+\Omega_i(x)^{2}})/2$. These are shown schematically in Fig.~\ref{fig:levelstructure}b).
The complete wavefunction can then be expanded in a basis of these adiabatic eigenstates, which play the role of Born-Oppenheimer channel functions,
\begin{equation}
|\Phi(t)\rangle=c_+ (x,t) |\Psi_{+}\rangle+c_- (x,t) |\Psi_{-}\rangle,
\end{equation}
resulting in the equation
\begin{equation}
i\hbar\frac{\partial}{\partial t}c_{\pm}(x,t)=[H_M+V^{\pm}(x)]c_{\pm}(x,t)+H_{M}^{\pm}c_{\mp}(x,t),
\end{equation}
where $H_{M}^{\pm}=\langle\Psi_{\pm}(x,t)|H_M|\Psi_{\mp}(x,t)\rangle$ gives the non-adiabatic couplings between the dressed states due to the motion of the atom. Provided these latter terms are small, the two equations decouple and the atoms remain in a single dressed state. In our case, there will be no non-adiabatic loss of atoms in this sense, provided that they are loaded into low energy states of the lower of the two adiabatic potentials ($V^-$). If atoms are loaded into the higher energy dressed potential, loss of atoms into the continuum states of the $V^-$ potential can occur, however this is exponentially suppressed as the separation between adiabatic potentials is increased, with the loss rate $\Gamma_l \sim \Gamma_0 \exp(-\delta_i/\omega)$ \cite{yi08}, where $\Gamma_0$ is a prefactor that we do not compute in detail here, and $\omega$ is the trap frequency in an individual site of the dressed potential.

\subsection{Dressed potentials for independent states}

When these dressed potentials are created independently for different $m_I$ levels, the result is that we can form two independent but almost identical potentials. These can, e.g., be shifted with respect to each other in a 2D plane using interferometrically stable methods, e.g., by adding path length to an interferometer arm in which the light is frequency-shifted in order to produce one of the trapping frequencies. As discussed above, this means of creating spin-dependent lattices has substantial advantages over spin-dependent lattices for Alkali atoms, where the lifetime is limited by the need to tune the lattice beams to a frequency in the middle of the fine structure splitting. Here, the lifetime will be controlled by the lifetime of the $^3$P$_0$ level (which is many seconds), or by off-resonant couplings to shorter lived states (but these will typically be many tens of nanometers detuned). In the presence of a second frequency (e.g., due to the laser creating the lattice for the second internal state), atoms can also be lost from the lower adiabatic potential, essentially being coupled out of the lattice into the continuum. However, due to the large momenta in the resulting state, this rate is suppressed exponentially in the ratio of the separation between manifolds in a Floquet basis and the trapping frequency in the lattice, $\Gamma_l\sim \tilde \Gamma_0 \exp(-\omega_{\rm diff} / \omega)$ \cite{yi08}, where $\tilde \Gamma_0$ is a prefactor \cite{yi08} $\omega_{\rm diff}$ is the frequency difference of the lattices for the two qubit states.
 If we operate in a field $\sim 5000$ G, then $\omega_{\rm diff}\sim 2\pi \times 550$kHz, and if we choose the Rabi Frequency $\Omega \sim 120$ kHz, then $\omega\sim 15$ kHz. We can also reduce the required field strength by choosing $m_I$ levels that are further separated (in $^{87}$Sr we can reduce the required field strength by a factor of 9 by choosing $m_l=-9/2$ and $m_l=+9/2$ as the two trapped states). 
 
\subsection{Combining resonant and off-resonant potentials}
\label{sec:combining}
\begin{figure}[tb]
\includegraphics[width=8.5cm]{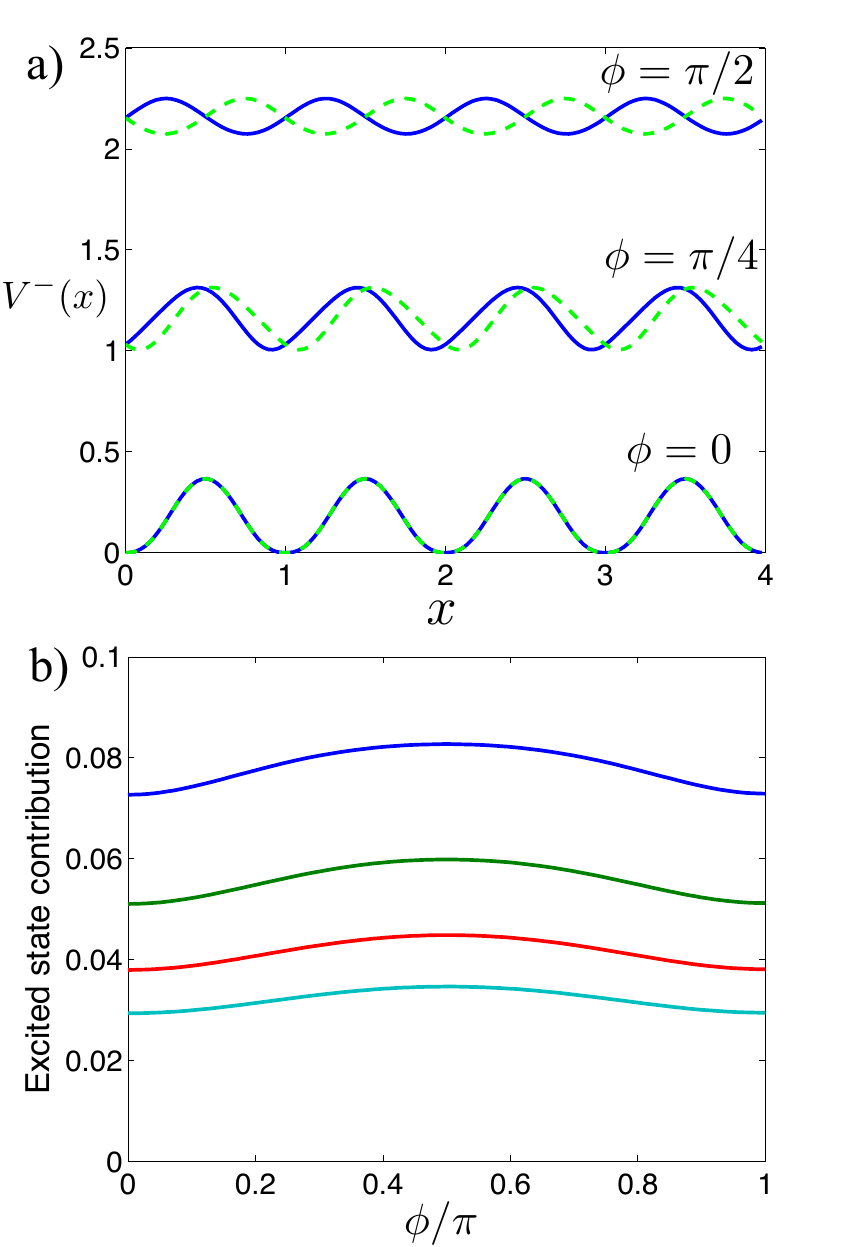}
\caption{a) Lower dressed potentials for the $\ket{0}$ qubit state (solid lines) and the $\ket{1}$ qubit state (dashed lines) formed by combining resonant and off-resonant contributions. These are plotted as a function of position for varying phase offsets $\phi$ between the coupling fields for the $\ket{0}$ and $\ket{1}$ states. For $^{87}$Sr at the wavelength of the clock transition, we obtain $\Delta E_{AC}^e \approx 3 \Delta E_{AC}^g$, and we choose the intensity so that $\Omega=4 \Delta E_{AC}^g$ (ca. 10 kW/cm$^2$). Here, $\delta=-3\Omega/4$. b) Projection on the excited state of the dressed state corresponding to the lower dressed potential, plotted for the same values of $\Omega$, $\Delta E_{AC}^{g}$, and $\Delta E_{AC}^{e}$ as in (a) as a function of $\phi$. The different lines correspond to varying $\delta$, from top to bottom, $\delta=-\Omega/2$, $-3\Omega/4$, $-\Omega$, and $-5\Omega/4$.} \label{fig:combinedpotentials}
\end{figure}

In practice, strong coupling at intensity $I$ on the clock transition will also give rise to off-resonant AC-Stark shifts $\Delta E_{AC}^e$ and $\Delta E_{AC}^g$ of the states $\ket{e,i}$ and $\ket{g,i}$ from coupling to other manifolds (e.g., $^1$P$_1$ and $^3$S$_1$) in addition to the resonant couplings between the two levels.  These must be added to the Hamiltonian, as $H=H_M+H_0+\Delta E_{AC}^e |e,i\rangle \langle e,i|+\Delta E_{AC}^g |g,i\rangle \langle g,i|$. As $\Delta E_{AC}^{e,g} \propto I$ and $\Omega\propto \sqrt{I}$, the off-resonant contributions will become more important as the intensity of the applied field becomes larger. For $^{87}$Sr, the shifts from the AC-Stark shift become of the same order as the AC-Stark splitting due to resonant coupling at relatively high fields, with $I \sim 50$kW/cm$^2$ \cite{Boydthesis, yi08}). At higher fields, the potentials $V^{\pm}$ will be modified by these shifts, but can still be made spin-dependent if the detunings and Rabi frequencies of the lattice beams are chosen carefully. This is illustrated in Fig.~\ref{fig:combinedpotentials}, where we show the lower adiabatic potential for each of the two nuclear spin states for a selection of different phases $\phi$ between the potentials. We see that at relative phase $\phi=0$ the potentials for different nuclear spin states are identical, and are given by a combination of the resonant and off-resonant contributions. At phase $\phi=\pi/2$, however, the off-resonant contributions from the two coupling frequencies, which are independent of the nuclear spin state, become spatially homogeneous due to the addition of the two spatially shifted contributions. At this point the sinusoidal form of the lattice potentials is due solely to the resonant contribution. It can be seen that the lattice, also in between, will be modified in such a way that the atoms will be transported through the lattice spin-dependently. 

We also note that two important characteristics relating to the shape of the lattice and the form of the dressed states changes as a function of $\phi$. Firstly, the lattice depth changes, because for $\phi=0$ the effect of the resonant and off-resonant contributions to the lower dressed potential are summed, whereas for $\phi=\pi/2$, the lattice is formed solely by a resonant contribution. This is shown in Fig.~\ref{fig:combinedpotentials}a. In addition, for $\phi=0$ the off-resonant potentials shift the coupling out of resonance, changing the adiabatic dressed states. As a result, the admixture of the excited internal state in the lower dressed level is relatively small. As the lattices are shifted, and the resonant contribution dominates, the admixture of the excited state increases. In Fig.~\ref{fig:combinedpotentials}b, we plot the admixture of the $^3$P$_0$ level, averaged over one period of the lower dressed potential for different values of the detuning $\delta$. We note that for the detuning values we choose here, this value is always small. This will lead to a significant reduction in collisional loss rates due to $^3$P$_0$-$^3$P$_0$ collisions when two atoms are on the same lattice site.

\subsection{Loading a quantum register}
In order to produce a quantum register with one atom in every lattice site, we begin from a spin-polarised gas of fermionic alkaline earth-like atoms, produced by optical pumping. This should be a degenerate Fermi gas so that the densities are sufficiently high to load a single atom per lattice site. Note that we choose Fermions here because for Yb and Sr, it is the fermionic isotopes that have non-zero nuclear spin, and thus allow us to encode qubits using this degree of freedom.
In the case that we have sufficient intensity to produce a large AC Stark shift at the same frequency as the final lattice, we can first load the gas into an off-resonance optical lattice in the $^1$S$_0$ state, and then adiabatically tune the coupling closer to resonance with the $^3$P$_0$ state in order to load the gas carefully into the lower dressed potential. A high-fidelity quantum register can then be formed by creating a  band-insulator state \cite{esslinger04}, and we gain substantially over the case where bosons would be used for a quantum register, as the temperature need only be substantially smaller than the bandgap, and not an interaction energy for the band insulator to form. In addition, if a harmonic trapping potential is added to the system, most defects in the state will be localised near the edges of the trap \cite{calarco04}. The resulting state can be further improved upon by applying additional techniques, such as filtering of the state to improve the fidelity \cite{rablloading} or fault-tolerant loading of atoms by transfer of atoms between two internal states, one trapped by the lattice and the other not \cite{agloading}.

\section{Single qubit addressing via the $^3$P$_2$ level}

\begin{figure}[tb]
\includegraphics[width=8.5cm]{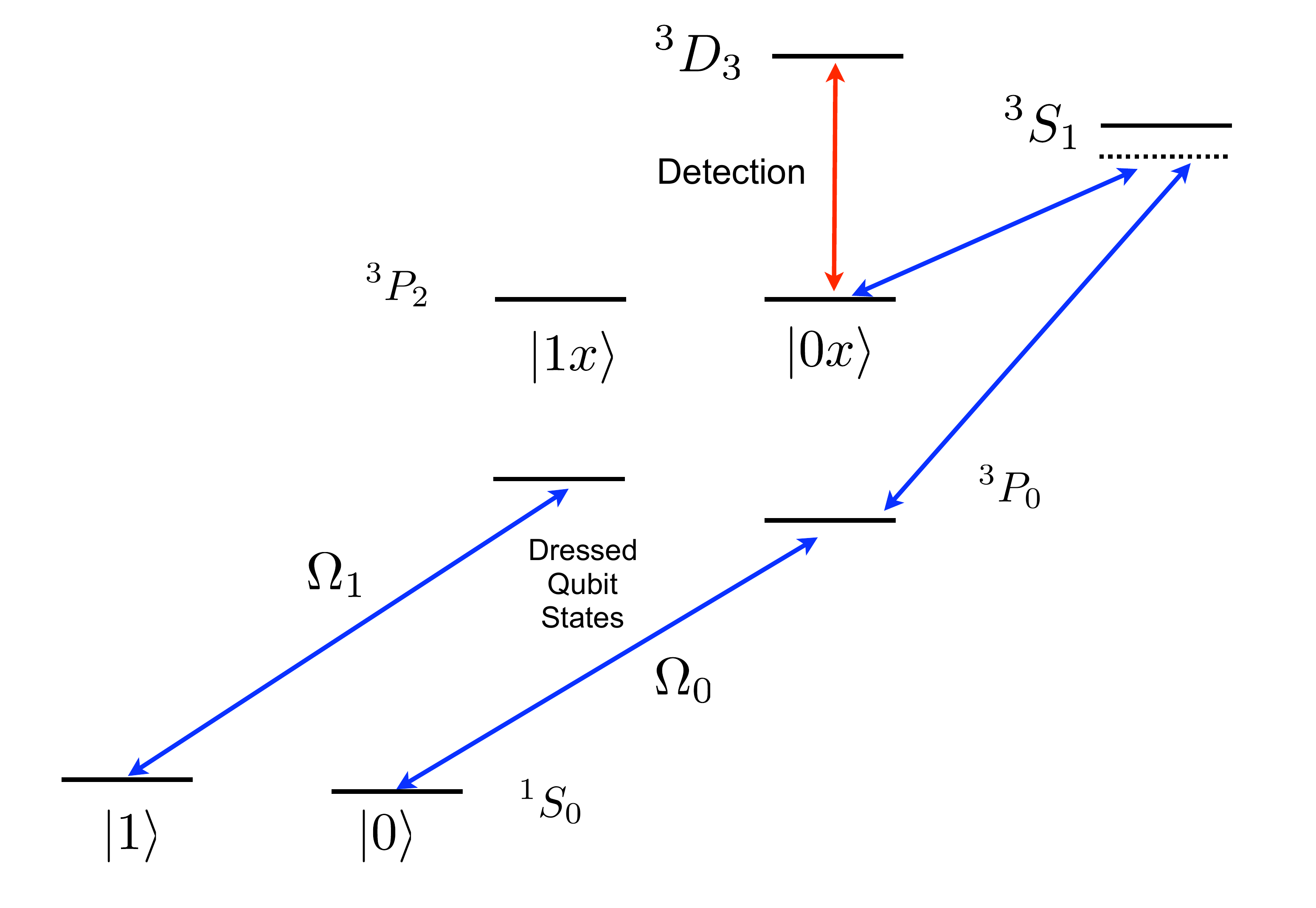}
\caption{Schematic diagram of qubit readout. Qubits are stored in dressed states $\ket{0}$ and $\ket{1}$, which are dressed superpositions of states in the $^1$S$_0$ and $^3$P$_0$ manifolds with a definite values of $m_I$. These can be coupled via off-resonant Raman processes to long-lived auxilliary states $\ket{0x}$ and $\ket{1x}$ in the $^3$P$_2$ manifold for the purpose of readout. In order to read out a particular qubit state, this state should be coupled to the $^3$P$_2$ manifold. It can then be detected by fluorescence on the cycling transition $^3$P$_2$-$^3$D$_3$.} \label{fig:readoutlevels}
\end{figure}

We would like to be able to read out the state of a single qubit, or alternatively perform gate operations on a single qubit. The has been enormous recent process in individual addressing of sites in an optical lattice via optical means \cite{addweiss,addchin,addgreiner,addgreiner2, addbloch, addbloch2,addott,addmeschede}. However, it would also be useful to be able to address individual qubits without the use of these techniques and the corresponding overheads in experiments. Such addressing can be achieved by coupling our dressed state qubit-selectively to states in the metastable $^3$P$_2$ level, and then detecting whether the atom is indeed present in the $^3$P$_2$ manifold. For the purpose of readout it is only necessary to be able to couple one of our two qubit states, e.g., the $\ket{0}$ state (which could be represented, e.g., by $m_I=-9/2$ in $^{87}$Sr) to an auxilliary level $\ket{0x}$ in the $^3$P$_2$ level (e.g., the $\ket{^3{\rm P}_2,F=13/2,m_F=-13/2}$ state, where $F$ is the total angular momentum quantum number $F$ and $m_F$ is the magnetic quantum number). The readout process is depicted schematically in Fig.~\ref{fig:readoutlevels}

Because of their non-zero electron spin, states in the long-lived $^3$P$_2$ manifold are much more sensitive to magnetic fields than the $^3$P$_0$ and $^1$S$_0$ level, and we can use these shifts to make possible a spatially-dependent readout of spin states by applying a magnetic gradient field, in a manner first mentioned in Ref.~\cite{dereviankoaddressing}. In applying such a field, the $^3$P$_2$ level can be significantly shifted, whilst the $^1$S$_0$ and $^3$P$_0$ states are not substantially shifted, and thus the form of the dressed lattice potential is not substantially changed. In particular, a gradient field of 1 G/cm will provide an energy gradient of 4.1 MHz/cm for the $\ket{^3{\rm P}_2,F=13/2,m_f=-13/2}$ state, or an energy difference of about 15 kHz between atoms in neighbouring sites for a field gradient of 100 Gauss/cm. Atoms in the dressed lattice can then be selectively transferred via a Raman process connecting off-resonantly via the $^3$S$_1$ manifold to the $^3$P$_2$ manifold, on a timescale limited by the frequency shift between neighbouring sites.

\begin{figure}[tbh]
\includegraphics[width=8.5cm]{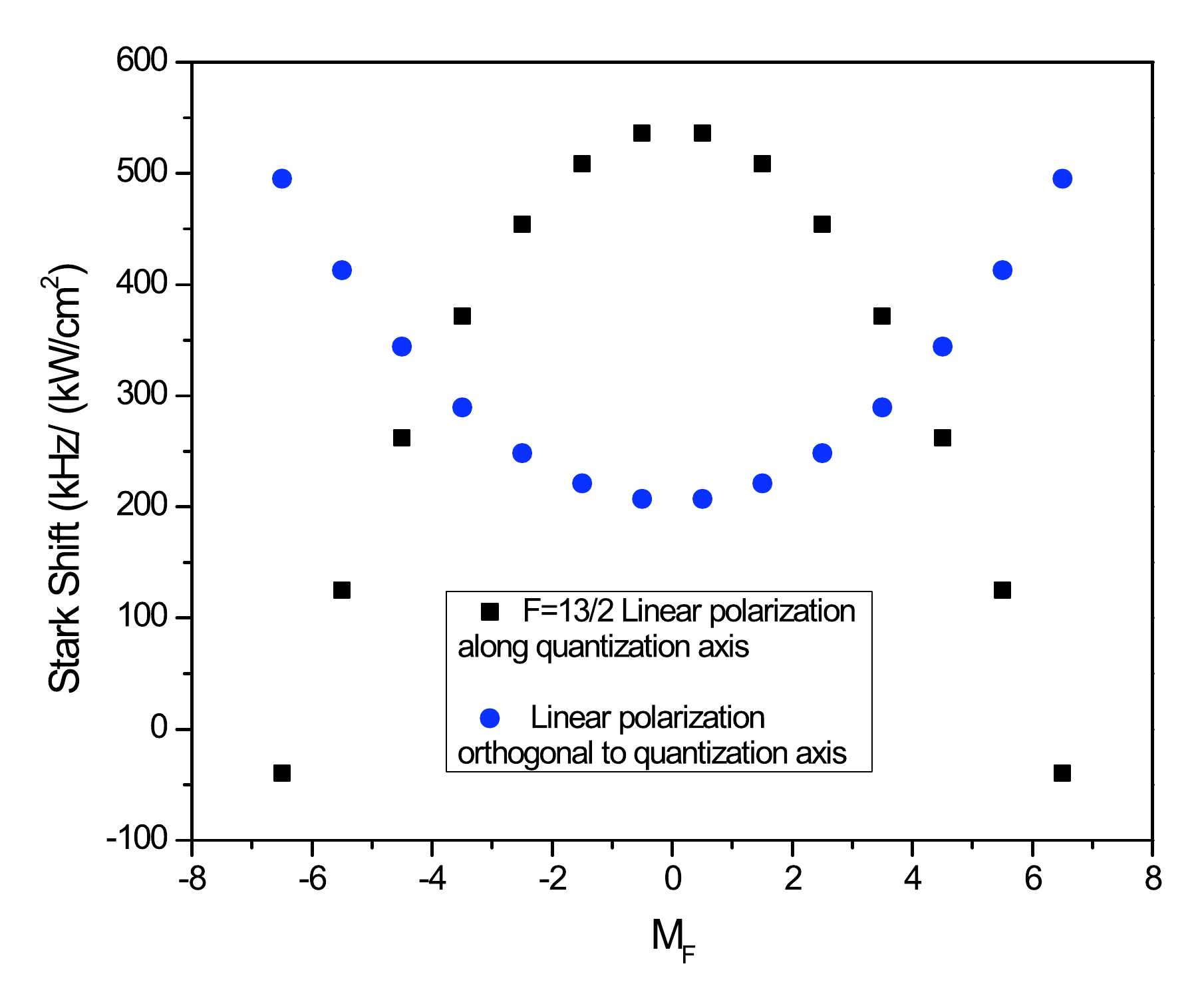}
\caption{AC-Polarisabilities for $^{87}$Sr in the $^3$P$_2$ manifold with $F=13/2$ at the frequency of the clock transition.} \label{fig:starkshifts}\label{fig:polarisability}
\end{figure}

This assumes, of course, that the state in the $^3$P$_2$ manifold to which we couple, $\ket{0x}$ is trapped in a lattice, preferably in a lattice at the same position as our qubit states $\ket{0}$ and $\ket{1}$. Thus, the most favourable states are those with  a significant negative $AC$-polarisability $\alpha$ at the wavelength of the clock transition, as the potential they experience due to the AC-Stark shift will have minima in the same places as the lower dressed state generated by the same lattice laser. We have computed the polarisability from known data of the states in the $^3$P$_2$ manifold of $^{87}$Sr, and have found that they vary substantially due to a large tensor shift. We write the shift $\Delta_E$ from linearly polarised light as
\begin{eqnarray}
h\Delta_E&=&-\frac{1}{2}\alpha E^2,\\
&=&-\left[\alpha^{\rm scalar} +\alpha^{\rm tensor}\frac{3m_F^2-F(F+1)}{F(2F-1)}\right]\frac{E^2}{2},
\end{eqnarray}
where we have separated the coefficients of the scalar and tensor shifts \cite{boyd07,Boydthesis}, and we obtain total polarisabilities at the clock transition frequency as shown in Fig.~\ref{fig:starkshifts}. Here we note that light polarised along the quantisation axis will give rise to a negative polarisability for the $F=13/2$, $m_F=-13/2$ state. This state is thus trapped by the same field creating the dressed lattice. We can couple from the $m_I=-9/2$ states in the dressed lattice via a Raman process directly into the
$F=13/2$, $m_F=-13/2$ state of the $^3$P$_2$ manifold, making this state ideal for use as the $\ket{0x}$ state in readout operations. A qubit could be read out by choosing the detuning of a Raman coupling between the $|^3$P$_2,\, F=13/2, \, m_F=-13/2\rangle$ state (auxiliary state $\ket{0x}$) and the 
$|-\rangle$ dressed state with $m_I=-9/2$ (qubit state $\ket{0}$) so that it is in resonance at only one site as a result of a gradient field shifting the energy of the
 $|^3$P$_2,\, F=13/2, \, m_F=-13/2\rangle$ state. Coupling of the $^3$P$_2$ level to a second qubit state with $m_I=-7/2$ would not occur as the $m_F=-11/2$ state is not trapped (if for a different species the equivalent state was trapped, then the large tensor shift would probably result in this transition being anyway out of resonance). The occupation of the $^3$P$_2$ level can then be determined by fluorescence measurements, e.g., using the cycling transition $^3$P$_2$-$^3$D$_3$, independent of the atoms remaining in the $^1$S$_0$ and $^3$P$_0$ levels. Note that the timescale for this readout process $\tau_{\rm readout}$ is limited by the trapping frequency in the dressed lattice potential, $\tau_{\rm readout} \gg 2\pi /\omega$. This requirement must be fulfilled so that the atom is not coupled to excited Bloch bands of the lattice. It is also desirable for this coupling to have similar trapping frequencies for the lattices trapping $\ket{0}$ and $\ket{0x}$, so maximising spatial overlap of the wavefunctions. Again, the $F=13/2$, $m_F=-13/2$ state of the $^3$P$_2$ manifold is favourable for this, as the polarisability indicates that the lattice depth will be around $150$ kHz for $I\sim3$kW/cm$^3$, which is a similar depth to that of the lattice for the dressed levels at the same lattice intensity (assuming that the detuning of the resonant coupling lasers, $\delta_i$ is small).
 
Note that one could equally use states with $|m_I|<-13/2$ in this process if one stores the qubit states in the upper dressed potential. This is disadvantageous, because a large detuning $\delta$ must be chosen for the lattice lasers to prevent non-adiabatic loss of atoms from the potential \cite{yi08}. Alternatively, an additional standing wave at a different frequency could be added to trap states in $^3$P$_2$ manifold via an additional AC-Stark shift. 

An alternative to using magnetic field gradients for addressing would be to apply a laser with spatially varying intensity at the magic wavelength (for equal shifts of the $^3$P$_0$ and $^1$S$_0$ levels. This would provide a position-dependent differential AC-Stark shift between the qubit states and the $^3$P$_2$ level, without affecting the relative energy of the $^3$P$_0$ and $^1$S$_0$ levels, and thus the dressed lattice.

\section{Quantum Gates in spin-dependent potentials}

Single-qubit gates can be performed in one of two ways in this scheme. The simplest means to obtain a global rotation of many qubits is to directly couple the dressed states for two nuclear spins via a Raman process. Alternatively, different nuclear spin (qubit) states can be alternately coupled to auxiliary states in the $^3$P$_2$ level in order to provide individual addressing for single-qubit rotations using the techniques described in the previous section. Such coupling requires the use of a trapped state in the $^3$P$_2$ manifold that can be coupled to both qubit states.  For $^{87}$Sr, such addressing for single-qubit operations would thus mean either using an auxiliary lattice to trap states from the $^3$P$_2$ manifold, or using the upper dressed states for qubit storage.

Two-qubit gates can, in principle, be performed similarly to exisiting schemes for alkali atoms, making use of the spin-dependent potentials. In particular, exisiting schemes for controlled collisions can be used to produce controlled-phase gates for atoms in neighbouring sites \cite{spinlattice1}. This has been implemented experimentally in a proof-of-principle experiment with alkali atoms \cite{spinlattice2}, but here we could take advantage of the 2D spin-dependent lattices without having to tune trapping lasers between fine-structure states.
  
\begin{figure}[tb]
\includegraphics[width=8.5cm]{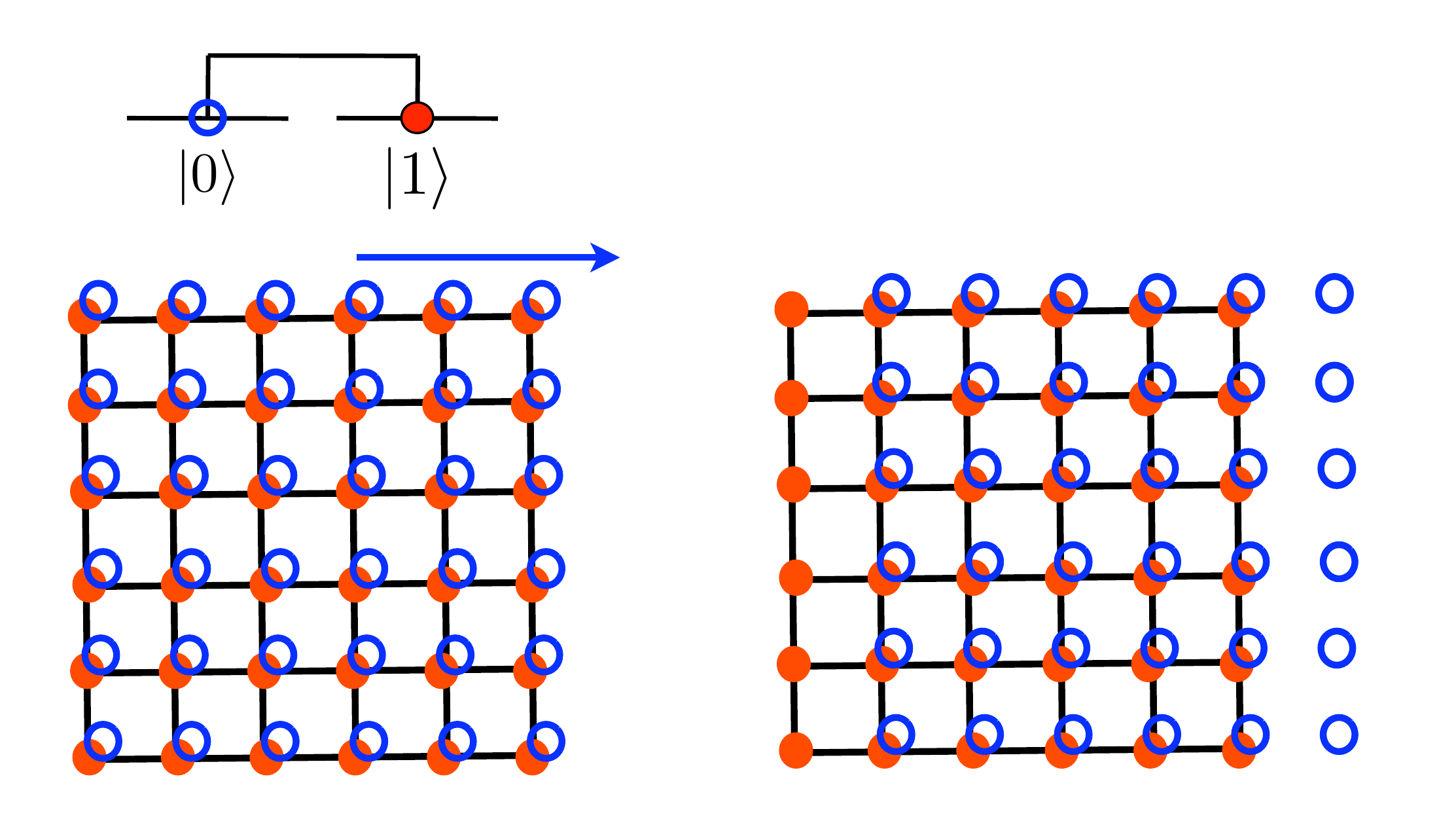}
\caption{If the qubits are trapped in a spin-dependent lattice, it is possible to shift the lattice for one qubit state by one site, so that neighbouring atoms are brought together only if the qubit to the left was in state $\ket{0}$ and the qubit to the right was in state $\ket{1}$. This can be used to aid in producing two-qubit quantum gates (see text for details).}\label{fig:spindep}
\end{figure}

These schemes can be seen to implement controlled-phase gates in three steps:
\begin{enumerate}
\item The spin-dependent lattices for each state are shifted relative to each other so that atoms at a chosen distance, e.g., in neighbouring lattice sites, will come together on the same site if and only if they were originally in a specific combination of qubit states. For example, if we write the state of a pair of neighbouring qubits as $\ket{q_1 q_2}$, where $q_1$ is the state of the first qubit and $q_2$ is the state of the second qubit in the pair, then atoms in the state $\ket{01}$ are brought together, whilst $\ket{00}$, $\ket{11}$ and $\ket{10}$ remain separated (see Fig.~\ref{fig:spindep}).
\item A phase shift is generated conditioned on whether two atoms are on the same site or not.
\item The atoms are returned to their initial positions.
\end{enumerate}
The phase in step two can be generated in a number of different methods, including via direct collisional phase shifts, or the use of blockade mechanisms. These different mechanisms are discussed in the following two subsections.
\subsection{Phase for two-qubit gates: controlled collisions}

For alkali atoms, the phase in step 2 is generated by collisional interactions between atoms. This could be performed directly if the atoms used have a relatively large scattering length in the $^1$S$_0$ manifold (e.g., $^{87}$Sr). For other species and isotopes such as $^{171}$Yb, this could also be achieved using optical Feshbach resonances \cite{srfeshbach,ybfeshbach} to enhance the otherwise very weak collisional interaction. The speed of such gates is limited by the strength of the on-site interaction between atoms, which for a single-band model is limited by the trap frequency in each lattice site, $\omega$.

However, the existence of weak collisional interactions for certain isotopes also motivates us to look at other gate schemes, particularly using excitations to states in the $^3$P$_2$ manifold.

\subsection{Use of $^3$P$_2$ levels}
We would again like to make use of states in the $^3$P$_2$ manifold to which our dressed qubit states (for a fixed nuclear spin) can be coupled, and which are trapped in the same locations as our qubits. This time we will assume that we have two such auxiliary levels, $\ket{0x}$ and $\ket{1x}$, as depicted in Fig.~\ref{fig:readoutlevels}. 

\subsubsection{Phase for two-qubit gates: dipole blockade mechanism}
\begin{figure}[tbh]
\includegraphics[width=8.5cm]{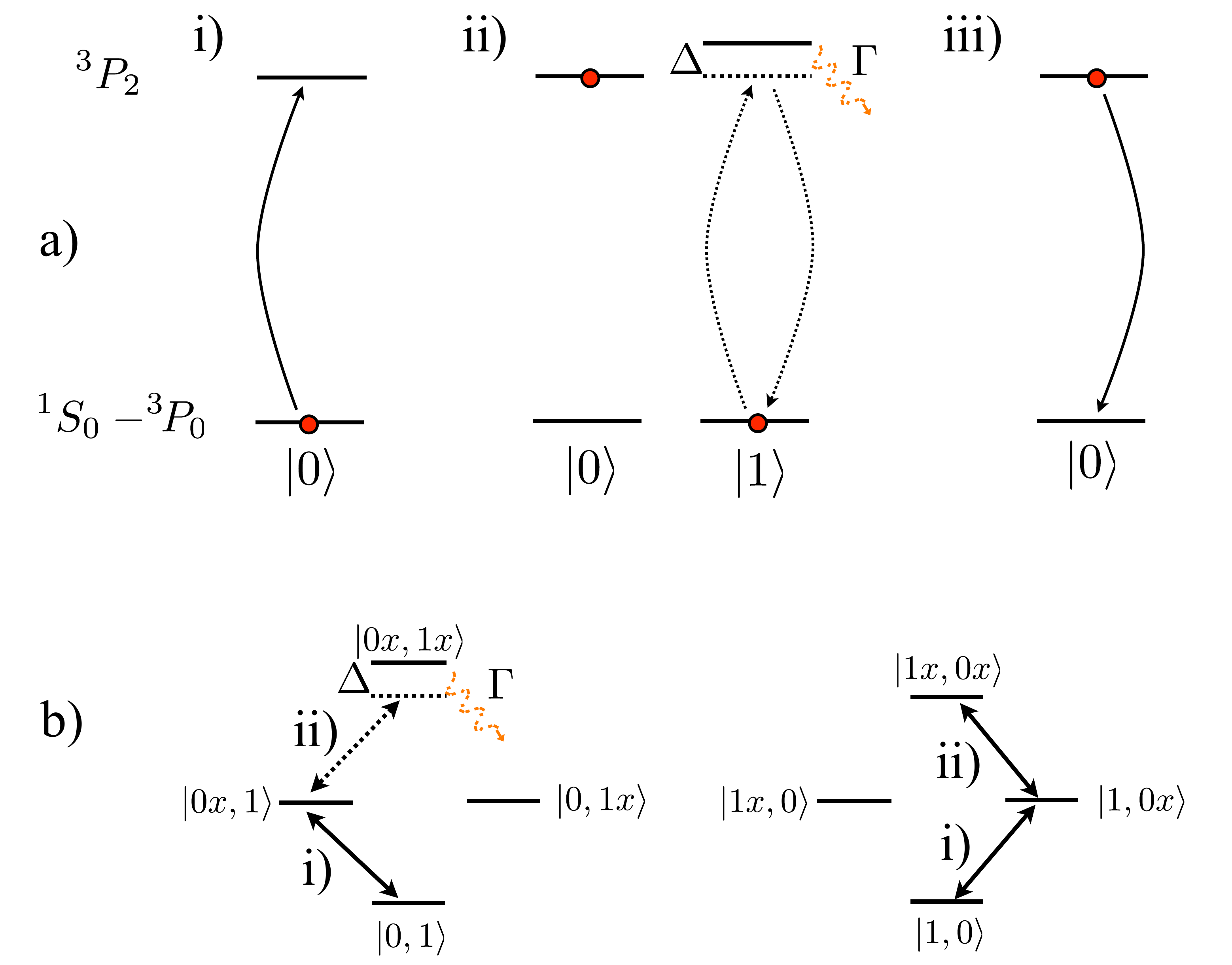}
\caption{Schematic diagram of a blockade gate including loss. (a) Operations performed on the individual qubit states $\ket{0}$ and $\ket{1}$. (see text for details) (b) Comparison of the operations for initial two-qubit states $\ket{0,1}$ and $\ket{1,0}$ in neighbouring qubits, showing the two-qubit levels. } \label{fig:lossyblockade}
\end{figure}

 For sufficiently large onsite dipole-dipole interactions, which provide a energy shift between $^3$P$_2$-$^3$P$_2$ collisional interactions and $^3$P$_0$/$^1$S$_0$-$^3$P$_2$ corresponding to a large frequency shift $\Delta$, we can use a dipole blockade mechanism to produce a $\pi$ phase shift, as proposed, e.g., for Rydberg atoms \cite{int:rydberg}. This is illustrated in Fig.~\ref{fig:lossyblockade}, and consists of 3 steps:
 \begin{enumerate}
 \item Excite all $\ket{0}$ qubit states to an auxillary level $\ket{0x}$ with a $\pi$-pulse
 \item Couple all $\ket{1}$ qubit states to an auxillary level $\ket{1x}$ with a $2\pi$-pulse at Rabi frequency $\Omega$, assuming that there is no collisional interaction between the $\ket{0x}$ state and either $\ket{1}$ or $\ket{1x}$ (i.e., the pulse duration $T$ is given by $\Omega T=2\pi$. In the ideal case, if the two atoms are on the same site (as will happen for an initial state $\ket{0,1}$, this step should be blocked by collisional interactions, which detune the coupling by a frequency $\Delta$.
 \item Return the $\ket{0x}$ state to the $\ket{0}$ state with a $\pi$ pulse.
 \end{enumerate}
 
Assuming there is no coupling of the qubit state $\ket{1}$ to the auxillary $\ket{1x}$ when the atom is on the same site as an already excited $\ket{0x}$ state (i.e., the blocking is perfect), the states of the two-qubit system after each step of this protocol are given in table \ref{table1}.
 \begin{table}[h]
 \begin{tabular}{|c|c|c|c|}
 \hline
Initial State & After Step 1 & After Step 2 & After Step 3 \\
\hline
 $\ket{0,0}$ &$ - \ket{0x,0x}$ & $- \ket{0x,0x}$ &$\ket{0,0}$\\
 $\ket{0,1}$ &$ -i \ket{0x,1}$ & $-i\ket{0x,1}$ &$-\ket{0,1}$\\
 $\ket{1,0}$ &$ -i \ket{1,0x}$ & $i \ket{1,0x}$ & $\ket{1,0}$\\
 $\ket{1,1}$ &$ \ket{1,1}$ &$-i\ket{1x,1x}$ &$\ket{1,1}$\\
 \hline
 \end{tabular}
 \caption{The state of a two-qubit system after each step of the protocol for a blockade gate.}
 \label{table1}
 \end{table}
 
In practice, the state $\ket{0,1}$ will collect a small additional phase $\phi\sim \Omega/\Delta$, where $\Omega$ is the coupling Rabi frequency and $\Delta$ the detuning from the excited state, generated by the difference between $^3$P$_0$.

\subsubsection{Lossy blockade mechanism}

It was shown by C. Greene and his collaborators \cite{greene1,greene2} that, in fact, two-body collisions of atoms in the $^3$P$_2$ level lead to large inelastic loss. However, this loss can actually help us in producing the blockade effect, as large losses involving coupling to the continuum at a rate $\Gamma$ from a given level can also dynamically suppress occupation of that level, as is well known from the physics of a two-level system. In the limit where $\Delta \ll \Gamma$, this would even produce a blockade gate based entirely on a lossy blockade mechanism. In this way we can turn an apparent problem into a feature of the system. Such ideas have also been proposed in the context of quantum simulation with cold atoms in optical lattices, where three-body losses can be used to prepare interesting many-body states via a similar mechanism \cite{daley3body,cirac3body}

The key characteristics of the inelastic loss processes that make this possible are:

\begin{itemize}
\item{The energy change in the inelastic collision is larger than the lattice depth, so that the energy carried away as kinetic energy is sufficient to couple the atoms into the continuum of motional states.}
\item{The length scale on which the physics of the inelastic collision takes place is smaller than the confinement length in a lattice site, so we do not expect the loss process to be substantially modified by the presence of the lattice}
\item{The rates for loss are large, and could reach of the order of $\Gamma=2\pi\times20$kHz for lattice densities up to $10^{16}$cm$^{-3}$.}

\end{itemize}

In the presence of loss, the basic physics of the second step of the protocol, as illustrated in Fig.~\ref{fig:lossyblockade} then reduces to a two level system, where the state with one atom in $^1$S$_0$ and one in $^3$P$_2$ playing the role of a lossless ``ground'' state and that with two atoms in $^3$P$_2$ the role of the lossy excited state. If we write these states as a spin-1/2 system, the Hamiltonian reduces to 
\begin{equation}
H=\frac{\Omega}{2} (\sigma^+ + \sigma^-) -\frac \Delta 2 \sigma^z 
\end{equation}
where $\sigma^+=|e\rangle\langle g|$, $\sigma^-=|g\rangle\langle e|$ and $\sigma^z=|e\rangle\langle e|-|g\rangle\langle g|$ are the usual spin operators for our two-level system with lossy excited state $|e\rangle$ and lossless ``ground'' state $|g\rangle$, $\Omega$ is the Rabi frequency for the coupling laser, and $\Delta$ is the effective detuning from the excited state, which can be induced by interaction between two atoms when they are both in the $^3$P$_2$ manifold. Including the loss, this system is described by the master equation
\begin{equation}
\dot \rho =-i [H,\rho] -\frac{\Gamma}{2} \left[\sigma^+ \sigma ^- \rho + \rho \sigma^+ \sigma^- - 2\sigma^- \rho \sigma^+  \right]. \label{mastereq}
\end{equation}
In the limit $\Delta, \Gamma \gg \Omega$ we can describe the time evolution of a system initially prepared in the ground state in perturbation theory, giving the probability that no decay has occurred at short times $t$ as
\begin{equation}
p={\rm e}^{-\Gamma_{\rm eff}t},
\end{equation}
with
\begin{equation}
\Gamma_{\rm eff}\approx \frac{\Omega^2}{4(\Delta^2+\Gamma^2/4)}\Gamma \approx \frac{\Omega^2}{\Gamma}
\end{equation}
in the limit that $\Gamma \ll \Delta$. For our lossy blockade gate this is the worst-case scenario for loss events. We immediately see that the ratio of the loss time to the gate time (determined by $\Omega$) is given by
$\tau_{\rm loss}/\tau_{\rm gate}=\Omega/\Gamma$. This will give the fidelity of the lossy blockade gate.

\begin{figure}[tbh]
\includegraphics[width=8.5cm]{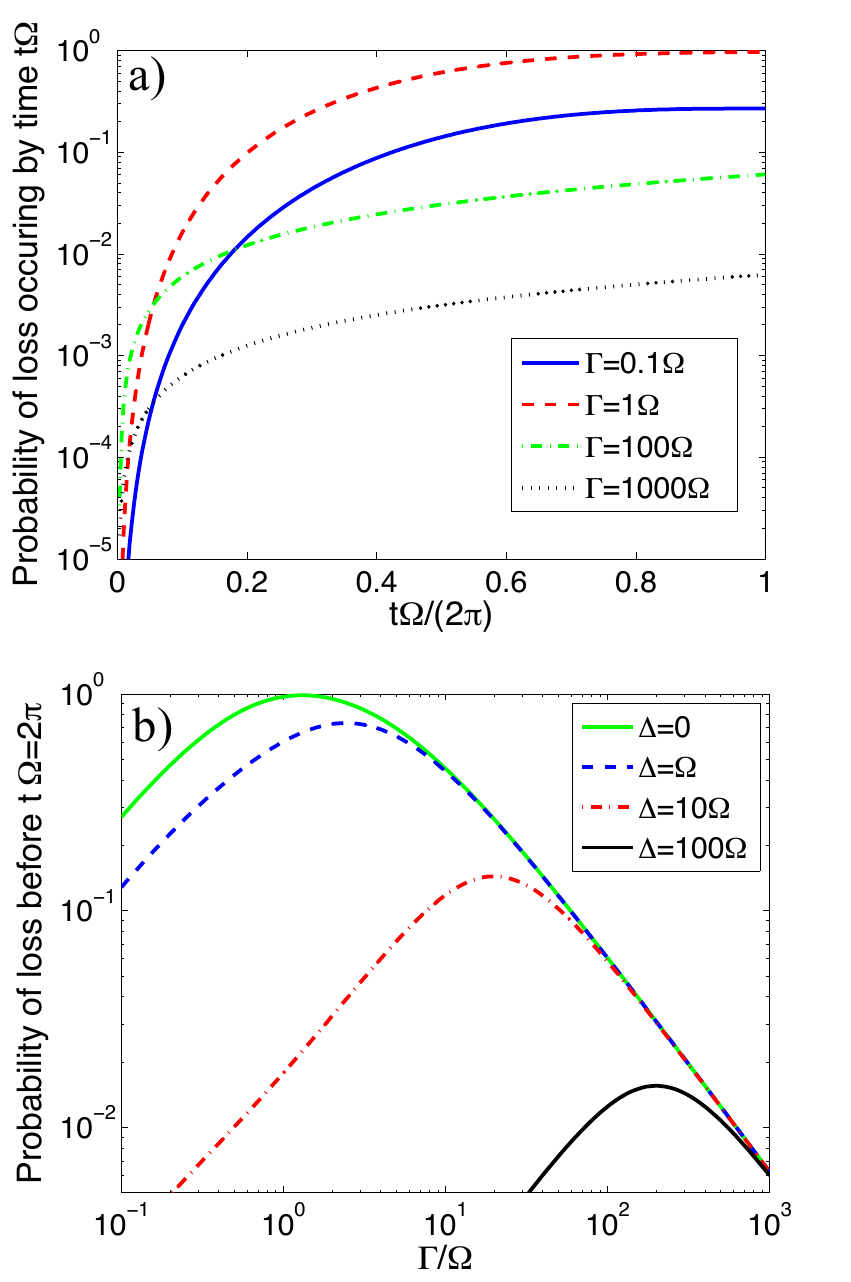}
\caption{Loss from a two-level system prepared in the stable state and coupled to the lossy state computed via integration of eq.~(\ref{mastereq}). (a) The probability that the system has undergone a loss event as a function of time when prepared in the ground state, with $\Delta=0$, for varying values of $\Gamma/\Omega$. (b) The probability that a system has undergone a loss event by time $t\Omega=2\pi$. These values represent the gate fidelity of a lossy blockade gate with $\Delta=0$, or with a combined blockade generated by interactions and loss with $\Delta\neq 0$. } \label{fig:twolevel}
\end{figure}

The blockade mechanism is illustrated in Fig.~\ref{fig:twolevel}, where we plot the decay probability as a function of time $t$, and then at fixed time $\Omega t =2\pi$ for varying $\Gamma/\Omega$.

\section{Other Two-qubit gates}
It would also be possible to make use of exchange interactions for fermions \cite{exchangegate}, but we will not discuss this in detail because it does not make specific use of the spin-dependent potentials, and does not, in its original form, take specific advantage of the properties of alkaline earth atoms.

Another possibility is the direct use of Rydberg gates \cite{int:rydberg}, which have been recently demonstrated for trapped alkali atoms \cite{Urban09,Gaetan09}. The separate hierarchy of Rydberg states for the singlet and triplet manifolds could give advantages for Rydberg excitations in alkaline earth atoms, especially facilitating easier state-dependent excitation. These could also be performed together with gradient addressing, exciting the Rydberg state from the $^3$P$_2$ manifold.

\section{Decoherence/loss Mechanisms}
There will be a number of possible sources of decoherence within this setup, all of which should be controllable in the experiment. These include magnetic field fluctuations, decoherence due to frequency noise of the lasers, spontaneous emissions, and collisional losses. We briefly summarise the role of these key sources of decoherence below.
\subsection{Magnetic field fluctuations}
Magnetic field fluctuations will probably still constitute the largest source of decoherence, however, this is reduced by almost 3 orders of magnitude compared with qubits encoded on an electron spin. Decoherence from local fluctuations in the magnetic field will contribute both due to direct shifts of the energy of the qubit states, and from the modification to the dressed potentials due to the differential shift between the $^1$S$_0$ and $^3$P$_0$ levels.

\subsection{Stability of the trapping lasers}
A finite laser linewidth for the dressing laser creating the potentials will give rise to fluctuations $\Delta \delta_i$ in the detuning $\delta_i$, and therefore the energy of atoms trapped in the dressed potential.
However, these fluctuations will lead to the same fluctuation in lattice depth for the two qubits, $\Delta \delta_0=\Delta \delta_1$ . The resulting ground state energy will shift by different amounts, as the lattice periods are different. However, if the corresponding wavelengths are $\lambda$ and $(1+\varepsilon)\lambda$, then the difference in trap frequencies for the two qubit states, $\Delta_\omega$ is given in terms of the depth fluctuations $\Delta V$ by
\begin{equation}
\frac{\Delta_\omega}{2} = \sqrt{\Delta V \frac{4\pi^2 \hbar^2}{2m\lambda^2} }-\sqrt{\frac{\Delta V 4\pi^2 \hbar^2}{2m\lambda^2 (1 + \varepsilon)^2} }\approx  -\varepsilon\sqrt{\delta V \frac{4\pi^2 \hbar^2}{2m\lambda^2}} 
\end{equation}
Thus, as $\varepsilon\sim 10^{-8}$, this decoherence mechanism will be strongly suppressed, and for laser linewidths of the order of tens of Hz, dephasing times can be many minutes. On the other hand, the resulting noise $\Delta V$ on the depth of the lattice could give rise to heating of the particles to higher oscillator levels, if appropriate frequency components are present in the noise in order to drive these coupling.s This would lead to imperfect couplings for gate and readout operations. Such heating rates can be estimated \cite{heating1,heating2} as giving an energy increase $\langle \dot E \rangle=\Gamma_{\rm heat} \langle E \rangle$ with rate $\Gamma_{\rm heat}= \pi^2 \omega^2 S_e(2\omega)/2$, where $S_e(2\omega)$ is the one-sided power spectrum of the trap amplitude noise at twice the trap frequency $\omega$. In our case, as for $\Omega_i \gg \delta_i$ $\Delta V \approx \Delta \delta_i^2/\Omega^2$, this is also suppressed by an extra factor of $\Delta \delta/\Omega$. In addition, other sources of heating, such as intensity noise on the lasers creating the lattice, or shaking of the lattice potential (due, e.g., to vibrating optical components) will have a similar effect \cite{heating1,heating2}. 

\subsection{Spontaneous emissions}
Qubits can decohere or be destroyed (the atoms lost from the lattice) by spontaneous emission events. These can come from two sources: the finite lifetime of the $^3$P$_0$ state, and off-resonant coupling to states with a short lifetime induced by the lattice lasers (storage) or coupling lasers (during gate and readout operations). However, the lifetime of $^3$P$_0$ is many seconds, and this source of atom loss can be further suppressed by using the resonant lattices only for spin-dependent transfer, and storing the atoms at other times in the $^1$S$_0$ state (see previous subsection).  
  
 \subsection{Collisional losses from $^3$P$_0$}
Measurements of collisional losses between atoms in the $^3$P$_0$ manifold are currently underway in several groups, in order to determine what the collisional lifetime is when two atoms are present in these states at the typical lattice densities that will be encountered here (ca. $10^{14}$cm$^{-3}$-$10^{15}$cm$^{-3}$ onsite). Effects of these losses have been observed recently, e.g., in samples of Strontium atoms confined in 1D tubes \cite{bishof2011}. However in our case, during storage, readout, and single-qubit operations, the atoms are anyway isolated by the lattice, and two atoms will not collide. Thus, the only time that two atoms with components of states from the $^3$P$_0$ manifold are present on the same site is during two-qubit gate operations. If these take place on a timescale ca. $1$ms, then we would require collisional stability of our atoms for timescales longer than $100$ms in order to achieve gate fidelities larger than $99$\% if both atoms were in the $^3$P$_0$ manifold. However, as shown in Sec.~\ref{sec:combining}, the combination of resonant and off-resonant lattices mean that the amplitude for atoms to be in the $^3$P$_0$ manifold is small for all stages of operation except during transport of atoms. Gate schemes can be made more immune to these losses by using larger intensity trapping lasers, and thus introducing a larger component from the off-resonant lattice (see Sec.~\ref{sec:combining}). This will ensure that when the lattices for the two qubit states overlap that the dressed states are dominated by off-resonant lattices for $^1$S$_0$, and that the admixture of the $^3$P$_0$ state is small. If the probability to find a single atom on a given site in the $^3$P$_0$ manifold is $\varepsilon_3$ for each of the qubit states, then the onsite loss rate will be suppressed by a factor $\sim \varepsilon_3^2$.
 
\section{Summary and outlook}

In summary, the quantum computing scheme we presented based on nuclear-spin-depenent lattices with near-resonant coupling on the clock transition for alkaline-earth(-like) atoms has several advantages over schemes with alkali atoms. The use of nuclear spins for qubit storage makes this scheme relatively robust against decoherence due to magnetic field fluctuations, and coupling to the $^3$P$_2$ manifold provides high-resolution individual qubit addressing with a magnetic gradient field. There are also possibilities here to perform gates based on transfer of states to long-lived metastable excited levels (e.g., $^3$P$_2$), including the new mechanism of lossy blockade gates. In comparison with a scheme presented previously using electronic-state-dependent lattices, this scheme does not require lasers that independently trap the $^1$S$_0$ and $^3$P$_0$ manifolds. This method is more sensitive to collisional losses between two atoms in the $^3$P$_0$ manifold, although this only affects short periods of time during the gate operations. While we have focused here on gates based on state-dependent lattices, other schemes, including Rydberg gates will benefit from the unique properties of alkaline-earth-like atoms. In particular, state-selective excitation to a Rydberg state would be simplified, e.g., by exciting one nuclear spin state to the $^3$P$_2$ manifold first.

The key experimental requirements for implemention of these methods are: (i) Large, stable magnetic fields (to provide the differential Zeeman shifts allowing spin-dependent lattices for different nuclear spin states; (ii) High-intensity stable laser on the clock transition (to provide a deep optical lattice whilst avoiding decoherence due to noise on the detuning $\delta_0$.); and (iii) Control over magnetic field gradients (to allow for either large parallel operations or individual addressing with qubits operations involving coupling to $^3$P$_2$ (although single-qubit gates could also be done directly in parallel, and the use of $^3$P$_2$ is only necessary in two-qubit gates in the case that the scattering lengths for the clock states are not sufficiently large).

Nuclear-spin-dependent lattices also have immediate possible application for quantum simulation with alkaline-earth-metal atoms. In particular, the dependence on the nuclear spin state could be used to break the degeneracy in models with SU(N) symmetry \cite{Cazalilla, Gorshkov}.

\begin{acknowledgements}
We thank Martin Boyd for close collaboration on the
previously presented scheme for quantum computing with alkaline earth atoms
(Ref.~\cite{aeshort}), and for estimates of polarizabilities used in
Fig.~\ref{fig:polarisability}. We thank L.-M. Duan, A. Gorshkov, C. Greene, and M. Lukin for stimulating discussions. J. Y. and P. Z. thank Caltech for hospitality during the course of this work, and A.J.D. thanks the Institute for Quantum Information at Caltech for hospitality. Work at JILA is supported by DARPA, NIST and NSF. Work in Innsbruck was supported by the Austrian Science Foundation (FWF) through SFB F40 FOQUS and EuroQUAM\_DQS (I118-N16) as part of the ESF EuroQUAM network, and also by the EU Networks NAMEQUAM and AQUTE.
\end{acknowledgements}

\end{document}